\documentclass[preprint,amsmath,amssymb,pre]{revtex4-1}
\usepackage{color,longtable}
\usepackage[dvips]{graphicx}
\usepackage{amssymb,amsfonts,amsmath}
\usepackage{comment}
\usepackage{here}

\begin{document}%
\title{Hidden Structural Control of Solvent Transport under Soft Jamming}
\author{Kento Tamaki$^{1}$}
\author{Naoya Yanagisawa$^{2}$}
\author{Rei Kurita$^{1}$}

\affiliation{%
1. Department of Physics, Tokyo Metropolitan University, 1-1 Minamioosawa, Hachiouji-shi, Tokyo 192-0397, Japan
}%
\affiliation{%
2. Komaba Institute for Science, Graduate School of Arts and Sciences, The University of Tokyo, Komaba 3-8-1, Meguro, Tokyo 153-8902, Japan
}%
\date{\today}

\begin{abstract}
Transport in soft jammed materials is often described as fluid motion through a fixed structure, leading naturally to capillary based descriptions.  
This picture appears particularly appropriate in strongly jammed systems, where structural rearrangements are suppressed and little visible motion is observed.
Here we investigate solvent transport in foam and show that this intuition fails to capture key aspects of the transport process.  
By directly observing both liquid penetration and bubble motion under controlled boundary conditions, we demonstrate that solvent transport is strongly influenced by the mechanical response of the foam structure, even though the intrinsic imbibition relative to the foam matrix remains purely capillary-driven. 
In closed systems, the jammed structure resists penetration and leads to a pronounced slowdown that cannot be accounted for by purely capillary descriptions.  
In contrast, in open systems, collective bubble motion accompanies solvent invasion, resulting in an apparent acceleration of transport.
These results indicate that the lack of structural motion does not guarantee a purely capillary description of transport.  
Our findings reveal a boundary controlled coupling between flow and structure, and highlight the need to reconsider transport processes in soft jammed systems, including foams, dense colloids, and biological tissues.
\end{abstract}

\maketitle

\section*{Introduction}
Soft jammed systems (SJS) are composed of densely packed deformable particles immersed in a solvent.
Such systems are found in a wide range of materials, including biological tissues, hydrogels, emulsions, and foams.
In contrast to classical hard-sphere jamming systems, the constituent particles in SJS can deform, allowing packing fractions that exceed the traditional jamming threshold and giving rise to persistent contacts between neighboring elements~\cite{Weaire2001,cantat2013}.

Solvent transport plays a central role in both the dynamics and function of SJS~\cite{cohen2013, Roveillo2020}.
In cell migration, pressure-driven solvent flow through crowded actin networks underlies protrusive motion~\cite{Charras2008}.
In colloidal dispersions, spontaneous flows induced by particle motion influence neighboring particles through hydrodynamic interactions, leading to collective aggregation and gel-like structures~\cite{Tanaka1997,Tanaka2000}.
A similar coupling has also been reported in scraping foams, where the onset of local rearrangements triggers a macroscopic transition from slipping to scraping motion~\cite{marchand2020, Endo2023}.
These examples demonstrate that solvent flow and microstructural dynamics cannot always be treated independently: the motion of one phase can strongly influence, or even enable, the motion of the other.
However, existing studies have primarily focused on situations in which structural rearrangements are present and directly observable.

In highly compressed SJS, where constituent particles form persistent contacts, the system develops excess interfacial energy that gives rise to a negative osmotic pressure.
Under equilibrium conditions, this pressure has been theoretically estimated and experimentally validated as a driving force for solvent uptake~\cite{Princen1986, Princen1987}.
Because structural rearrangements are strongly suppressed in such jammed states, solvent transport distance $x$ has often been described using capillary-based approximations, most notably the Lucas-Washburn relation, $x^2 \approx \frac{P r^2}{\eta} t$, where $P$ and $r$ denotes the driving pressure and the effective capillary radius.
However, several experimental studies have reported systematic deviations from the classical $t^{1/2}$ scaling, even in strongly jammed systems where particle rearrangements are negligible~\cite{cohen2013, Lorenceau2009, Mensire2015,Tsuritani2021}.  
In foam drainage experiments, for example, the effective osmotic pressure has been found to be significantly smaller than theoretical predictions when bubble motion is absent~\cite{Kaneda2025}.
These observations indicate that, despite the lack of visible structural motion, the jammed microstructure can still influence solvent flow.
The physical origin of this influence, and its quantitative impact on transport dynamics, remain unresolved.

To date, flow-structure coupling in soft jammed systems has been discussed mainly at a phenomenological level.
While experiments have suggested that structural rigidity can influence solvent transport, how such coupling operates when structural rearrangements are suppressed has remained unclear.
In this study, we address this issue by experimentally controlling boundary conditions in a model foam system and directly observing both solvent penetration and bubble motion.
We show that solvent transport is governed by the mechanical response of the jammed structure, even in the absence of observable rearrangements, and that boundary conditions determine whether this coupling accelerates or retards apparent transport.
By incorporating structural response into the force balance governing solvent flow, our results provide a quantitative framework that links solvent transport to structural mobility in soft jammed systems.

\section*{Results}
\subsection*{Experimental setup}
\begin{figure}[htbp]
\begin{center}
\includegraphics[width=150mm]{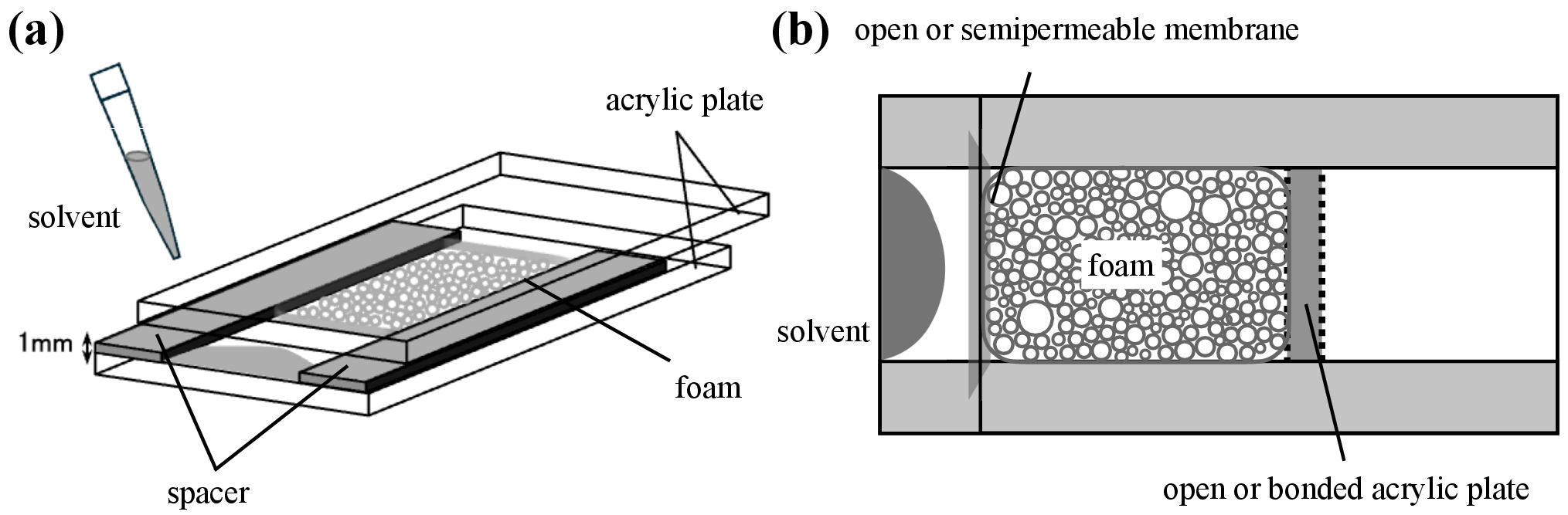}
\end{center}
\caption{\textbf{Experimental setup.} (a) Schematic diagonal view of the foam imbibition cell. A foam layer is confined between two acrylic plates with a spacer of thickness 1.0 mm. A dyed surfactant solution is brought into contact with one side of the foam.
(b) Top view showing the boundary conditions. The injection side is either open or covered by a semipermeable membrane that blocks bubble motion while allowing solvent transport. The opposite side is either open or sealed with an acrylic plate, defining fully open, half-open, and closed geometries.
}
\label{setup}
\end{figure}

We briefly describe the experimental setup shown in Fig.~\ref{setup};
full experimental details are provided in the Materials and Methods section.
A foam layer with a mean bubble diameter of 0.15~mm or 0.22~mm was confined between two acrylic plates using a spacer of thickness 1.0~mm.
The foam occupied a rectangular region with a height of 40~mm and a width of 30~mm, corresponding to a total foam volume of approximately $1200~\mathrm{mm}^3$.
At time $t = 0$, a volume of 800~$\mu$L of the same surfactant solution, dyed blue for visualization, was gently brought into contact with one side of the foam, and the subsequent solvent uptake dynamics were monitored.
To examine the role of boundary conditions, three configurations were investigated.  
In the full open system, both ends of the foam were exposed to air.  
In the half open system, a semipermeable membrane (mesh size 0.07-0.10 mm) was placed on the injection side, preventing bubble motion across the boundary while allowing solvent transport.  
In the closed system, the opposite side was additionally sealed with an acrylic plate, fully constraining bubble motion.

\subsection*{Boundary condition dependent imbibition dynamics}
We first examine how the imbibition dynamics depend on boundary conditions (see supplementary movies).
Figure~\ref{Dynamics}(a) shows the time evolution of the liquid penetration distance $x_{\mathrm{liq}}$
for a mean bubble diameter $d = 0.22$~mm and a liquid fraction $\phi = 0.10$.
Open circles, squares, and triangles represent $x_{\mathrm{liq}}$ measured under fully open, half-open,
and closed boundary conditions, respectively.
In all cases, the penetration dynamics can be described by a power-law form,
$x_{\mathrm{liq}} \sim t^{\alpha_1}$; however, the exponent $\alpha_1$ depends strongly on the boundary condition.

The fitted values of $\alpha_1$ are summarized in Fig.~\ref{Dynamics}(b) as a function of $\phi$.
At $\phi = 0.10$, for example, we obtain $\alpha_1 = 0.58$, $0.68$, and $0.37$ for the fully open,
half-open, and closed boundary conditions, respectively.
Previous studies have reported $\alpha_1 < 0.5$ even under fully open boundary conditions~\cite{cohen2013, Mensire2015,Tsuritani2021}.
The origin of this apparent slowdown will be discussed later.
By contrast, we observe $\alpha_1 > 0.5$, and notably a much larger exponent in the half-open geometry.
Such accelerated imbibition dynamics have not been reported previously.

\begin{figure}[htbp]
\begin{center}
\includegraphics[width=150mm]{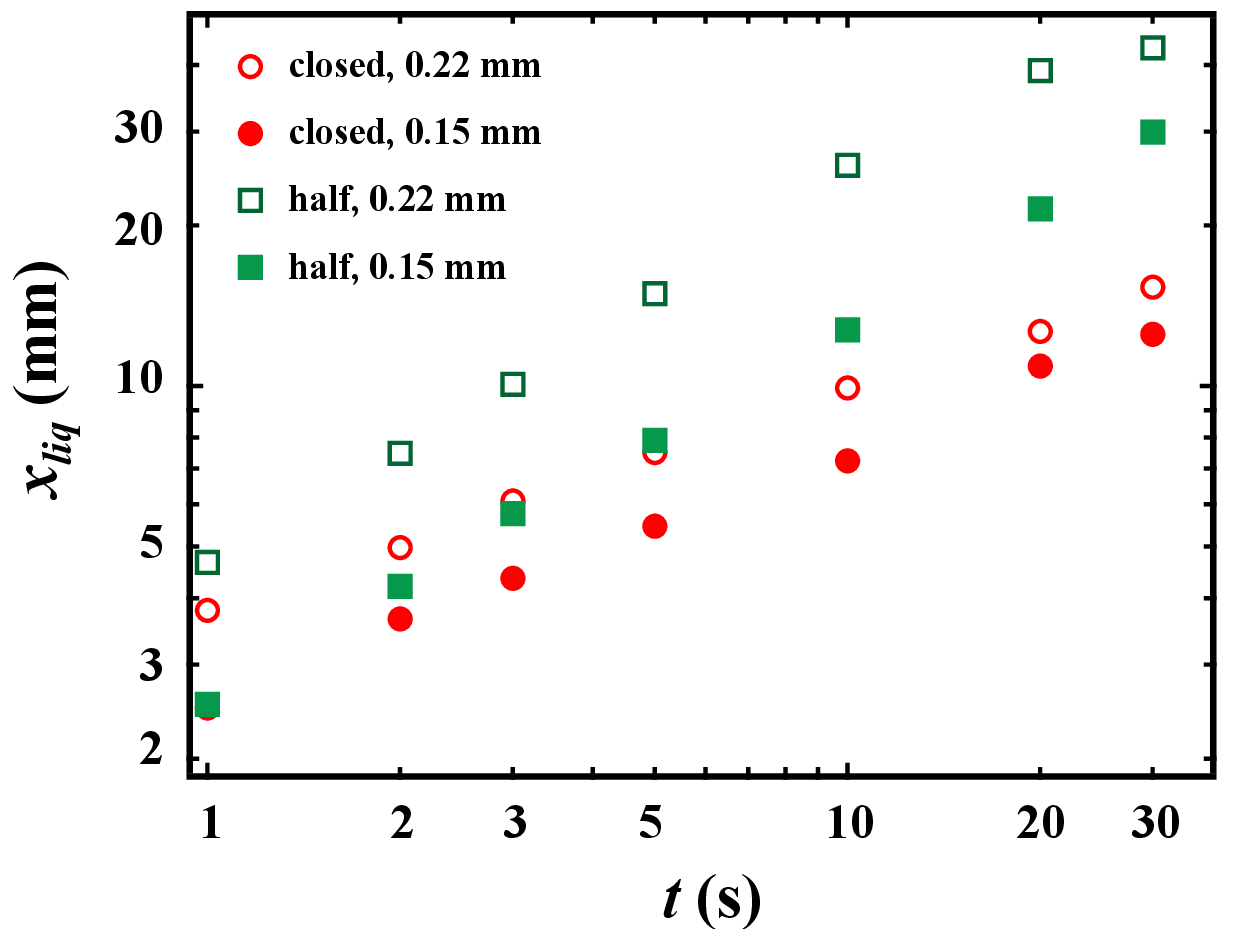}
\end{center}
\caption{
\textbf{Size-dependent imbibition dynamics.}
Time evolution of the liquid penetration distance $x_{\mathrm{liq}}$ for foams with mean bubble diameters $d = 0.22$~mm (opened) and $0.15$~mm (filled). Data are shown for closed (circles) and half-open (squares) boundary conditions. The foam composed of larger bubbles exhibits faster imbibition.
}
\label{size}
\end{figure}

This trend persists when the bubble size is varied.
Figure~\ref{size} shows the imbibition dynamics for $d = 0.15$~mm and $0.22$~mm at $\phi = 0.10$
under half-open and closed boundary conditions.
The exponent $\alpha_1$ remains nearly unchanged for different $d$,
indicating super-diffusive behavior in the half-open geometry and sub-diffusive behavior in the closed geometry.
In contrast, the imbibition speed itself depends on the bubble size, with larger bubbles exhibiting faster penetration.
This size dependence is opposite to common empirical expectations in industrial foam applications, and therefore has important implications for foam design.
As will be discussed later, however, this trend is physically natural within the present framework.
For clarity, the fully open boundary condition is not shown in Fig.~\ref{size}, as it would overcrowd the figure; qualitatively similar results are obtained.

\begin{figure}[htbp]
\begin{center}
\includegraphics[width=150mm]{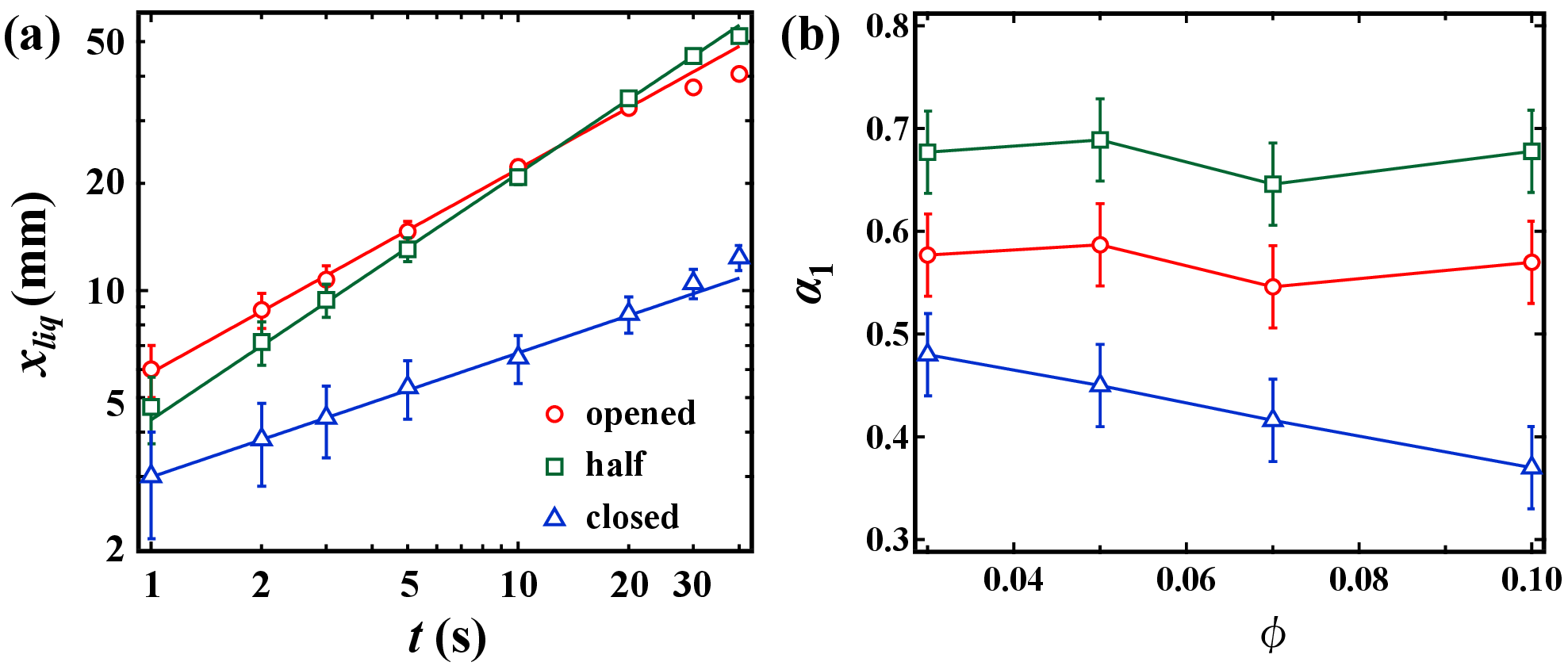}
\end{center}
\caption{
\textbf{Boundary-condition-dependent imbibition dynamics.}
(a) Time evolution of the liquid penetration distance $x_{\mathrm{liq}}$ for a foam with mean bubble diameter $d = 0.22$~mm and liquid fraction $\phi = 0.10$. Data are shown for fully open (circles), half-open (squares), and closed (triangles) boundary conditions. Solid lines are power-law fits.
(b) Imbibition exponent $\alpha_1$, defined by $x_{\mathrm{liq}} \sim t^{\alpha_1}$, as a function of the liquid fraction $\phi$ for different boundary conditions.
}
\label{Dynamics}
\end{figure}

\subsection*{Imbibition dynamics in the half-open boundary condition}
\begin{figure}[htbp]
\begin{center}
\includegraphics[width=150mm]{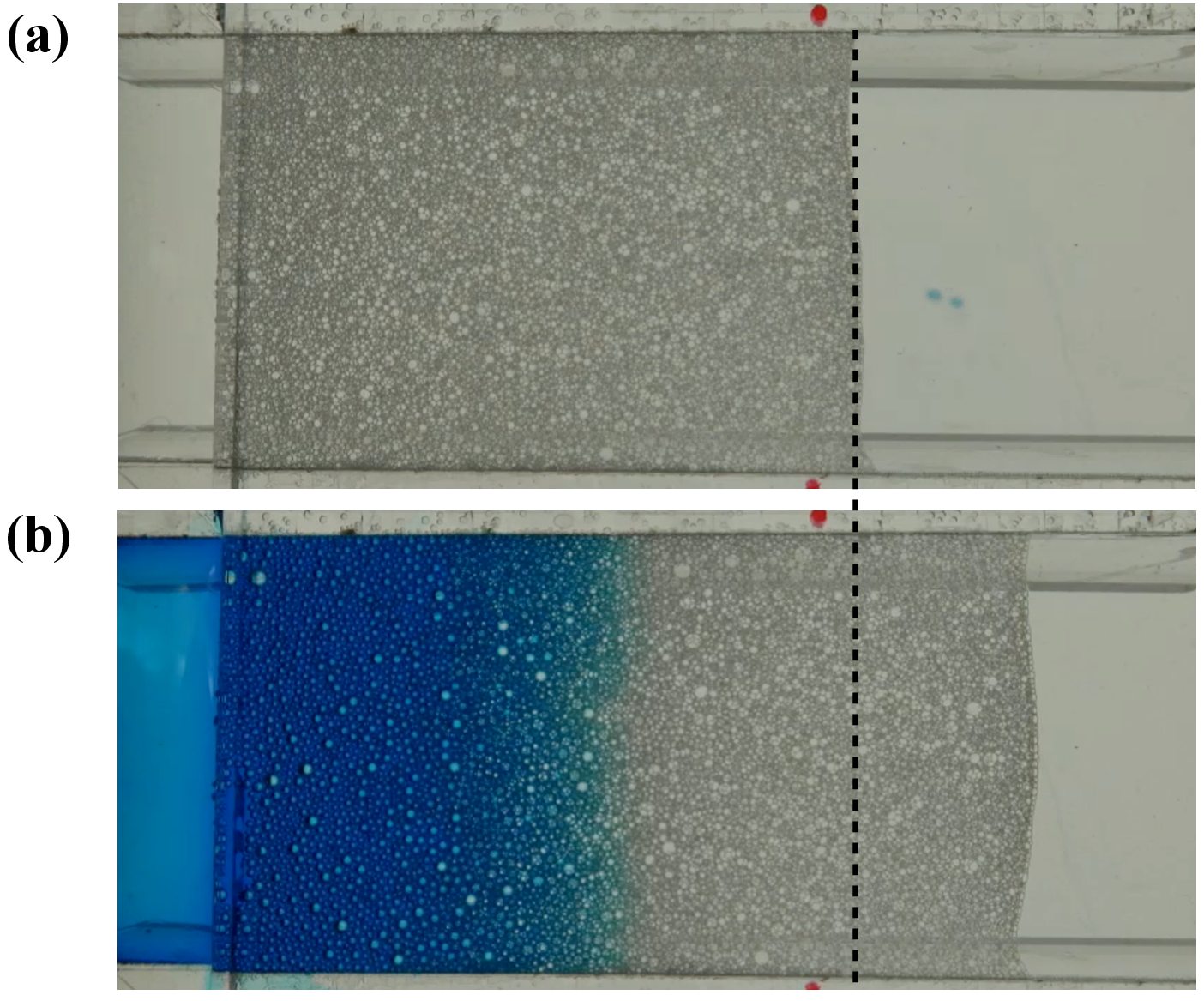}
\end{center}
\caption{
\textbf{Visual observation of foam motion during imbibition in the half-open geometry.}
Representative images of the foam before (a) and during (b) liquid imbibition. The dyed liquid penetrates from the left. The dashed line indicates the initial position of the right edge of the foam. As imbibition proceeds, the foam matrix translates in the direction of liquid invasion.
}
\label{half-image}
\end{figure}

We next focus on the imbibition dynamics in the half-open boundary condition.
Figure~\ref{half-image} shows representative images of a foam into which a blue-colored liquid is imbibed.
The left edge of the foam is confined by a semipermeable membrane, and the dotted line indicates the initial position of the right foam edge.
Figure~\ref{half-image}(b) shows the configuration at $t = 12$~s after the liquid is brought into contact with the foam.
As the liquid penetrates, the right edge of the foam is observed to move simultaneously with the liquid front.

\begin{figure}[htbp]
\begin{center}
\includegraphics[width=150mm]{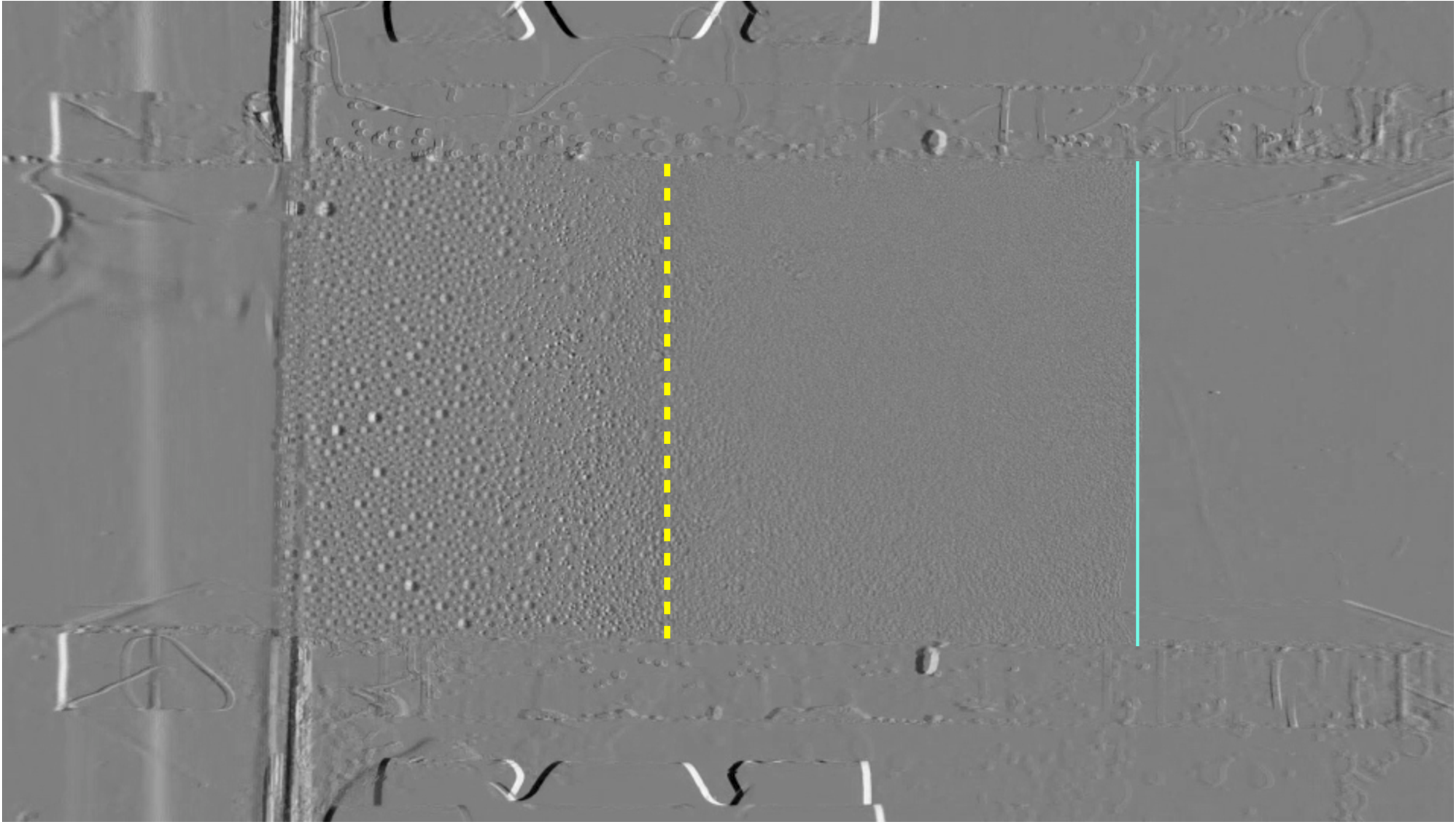}
\end{center}
\caption{
\textbf{Differential image analysis of foam motion.}
Differential image obtained from intensity fields over a time interval $\delta t = 0.5$~s during imbibition in the half-open geometry. The solid line marks the position of the right edge of the foam matrix, and the dashed line indicates the liquid penetration front. Heterogeneous intensity variations appear only within the imbibed region, whereas the foam ahead of the front undergoes an almost rigid translation.
}

\label{PIV}
\end{figure}

To examine whether this motion is accompanied by internal structural relaxation of the foam,
we perform image analysis on the recorded intensity fields.
From the image intensity $I(x,t)$, we compute the differential image $I_{\mathrm{dif}}(x,t) = I(x-\delta x, t+\delta t) - I(x,t)$, 
where $\delta x$ represents the displacement of the right foam edge during the time interval $\delta t$.
Figure~\ref{PIV} shows the resulting differential image for the time window $t = 12$--$12.5$~s,
with $\delta t = 0.5$~s and $\delta x = \text{0.05}~\mathrm{mm}$.
The solid line indicates the position of the right foam edge, which is used to determine $\delta x$, while the dashed line marks the liquid penetration front at $t = 12$~s.

In the region already invaded by the liquid, spatially heterogeneous intensity variations are observed,
indicating local bubble rearrangements during the time interval $\delta t$.
In contrast, ahead of the liquid penetration front, the intensity remains nearly uniform.
This uniformity indicates that bubbles in this region undergo a nearly rigid translation by $\delta x$
without internal rearrangements.
These observations demonstrate that, while the foam is locally deformed and rearranged within the imbibed region,
the foam matrix ahead of the front translates collectively as a whole.
A similar behavior is also observed under fully open boundary conditions.
In the fully open geometry, where both sides of the foam are open,
the foam matrix is found to move not only toward the non-injection side
but also toward the injection side (see Fig.S1).
We therefore interpret the observed motion as follows.
The imbibition flow compresses the foam matrix, generating elastic stresses,
which drive a collective translation of the foam toward the available open boundaries.

\subsection*{Imbibition dynamics in foam}
\begin{figure}[htbp]
\begin{center}
\includegraphics[width=150mm]{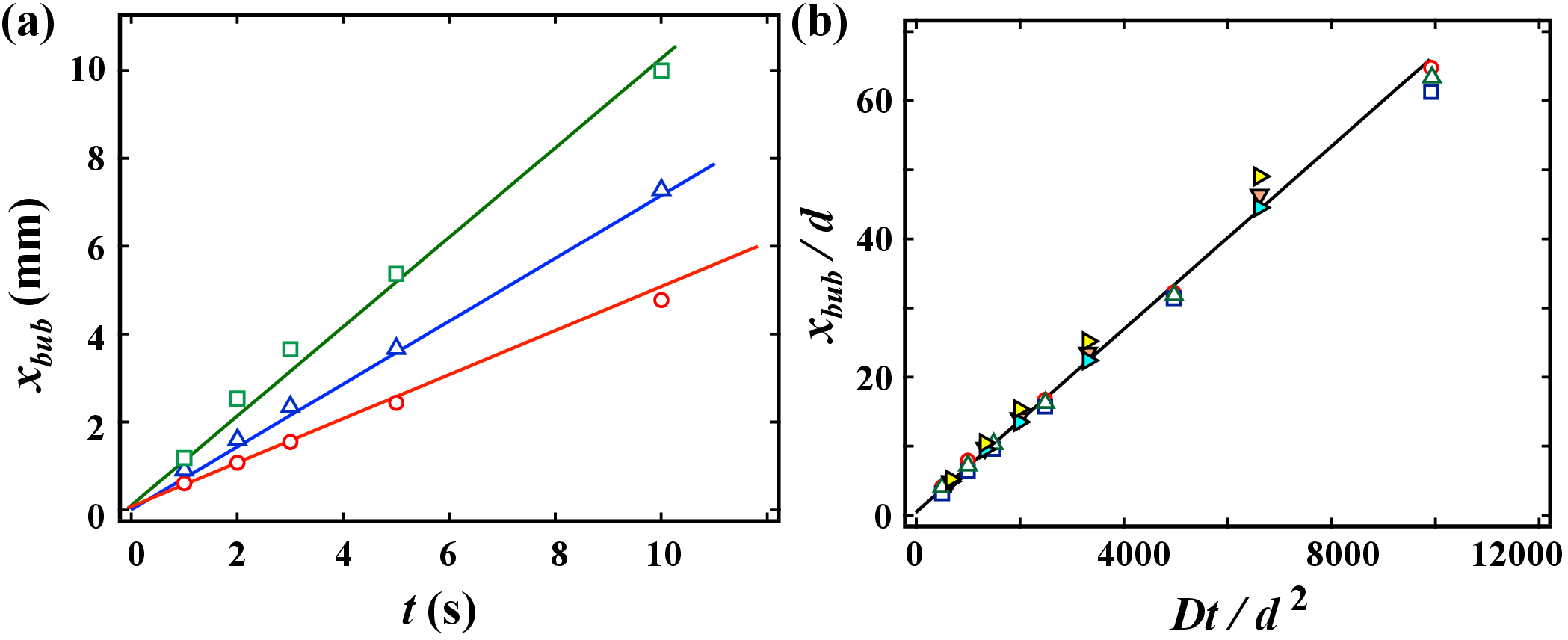}
\end{center}
\caption{
\textbf{Translation of the foam matrix during imbibition.}
(a) Time evolution of the foam matrix displacement $x_{\mathrm{bub}}$ obtained from edge tracking at $\phi = 0.10$. 
Symbols denote fully open (circles) and half-open boundary conditions with $d = 0.22$~mm (squares) and $d = 0.15$~mm (triangles). 
(b) Same data rescaled by the mean bubble diameter $d$. 
Filled and open symbols correspond to $\phi = 0.05$ and $\phi = 0.10$, respectively. 
Time is rescaled by $d^2/D$, where $D$ is obtained from the relative imbibition dynamics. 
The data collapse indicates an approximately constant matrix velocity over the observation time.
}
\label{bubble}
\end{figure}

\begin{figure}[htbp]
\begin{center}
\includegraphics[width=150mm]{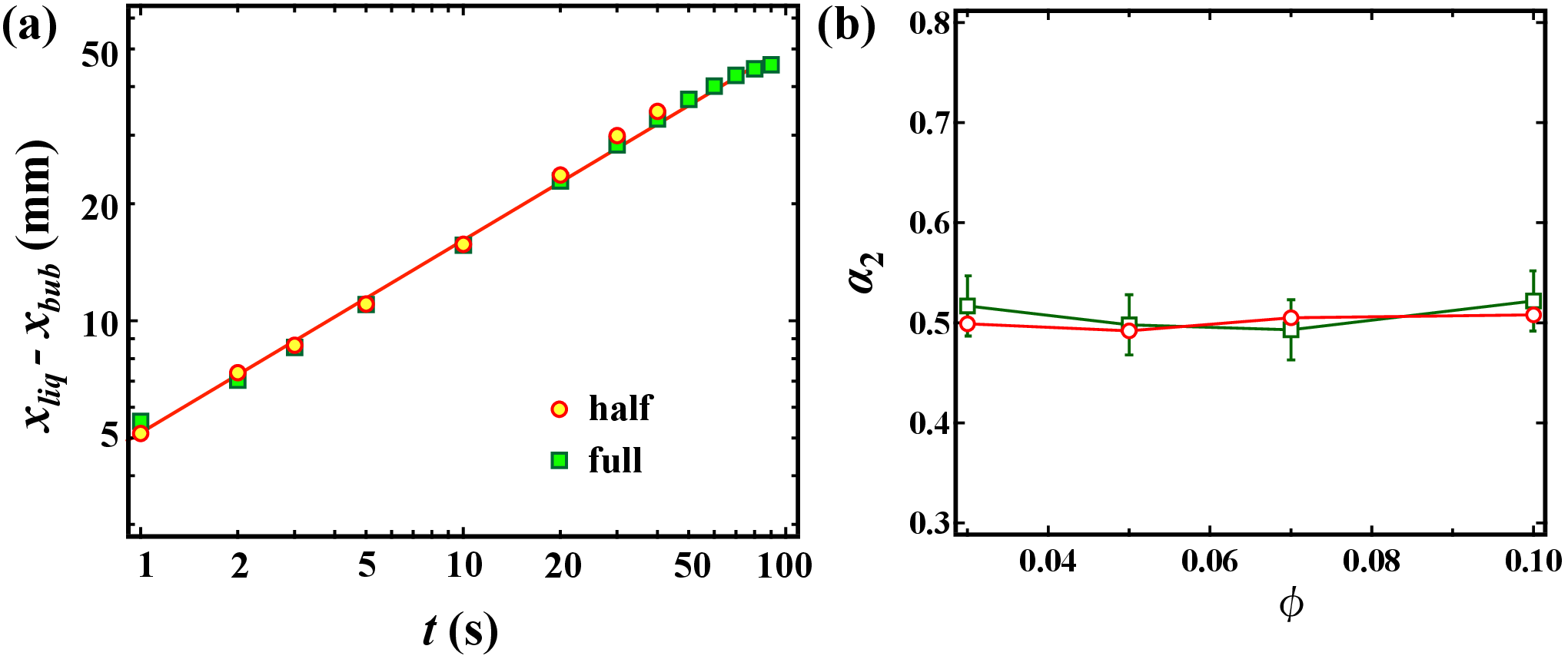}
\end{center}
\caption{
\textbf{Intrinsic imbibition dynamics relative to the foam matrix.}
(a) Relative imbibition distance $\Delta x = x_{\mathrm{liq}} - x_{\mathrm{bub}}$ as a function of time for half-open (circles) and fully open (squares) boundary conditions at $\phi = 0.10$. The solid line indicates the $t^{1/2}$ scaling.
(b) Exponent $\alpha_2$, defined by $\Delta x \sim t^{\alpha_2}$, as a function of the liquid fraction $\phi$. The exponent remains close to $1/2$, consistent with classical Lucas-Washburn imbibition.
}
\label{relative}
\end{figure}

We next examine how the right edge of the foam matrix, denoted by $x_{\mathrm{bub}}$, moves in response to liquid imbibition.
Figure~\ref{bubble}(a) shows the time evolution of $x_{\mathrm{bub}}$ at a liquid fraction $\phi = 0.10$.
Circles, squares, and triangles represent $x_{\mathrm{bub}}$ measured for $d = 0.22$~mm in the fully open boundary condition, and for $d = 0.22$~mm and $0.15$~mm in the half-open boundary condition, respectively.
In all cases, $x_{\mathrm{bub}}$ grows approximately linearly with time, indicating that the foam matrix translates at an almost constant velocity.
As discussed later, this constant-velocity translation can be rationalized by the balance between flow-induced driving and frictional resistance of the foam matrix.
The above observations indicate that, while the foam matrix undergoes a collective translation,
the Plateau borders ahead of the liquid front remain structurally intact.
We therefore consider the relative imbibition of the liquid with respect to the moving foam matrix,
defined as $\Delta x = x_{\mathrm{liq}} - x_{\mathrm{bub}}$.
Figure~\ref{relative}(a) shows the time evolution of $\Delta x$
under fully open (squares) and half-open (circles) boundary conditions at $\phi = 0.10$.
In both cases, $\Delta x$ follows a power-law scaling, $\Delta x \propto t^{\alpha_2}$,
with an exponent $\alpha_2 = 0.50 \pm 0.03$.
This scaling is found to be universal over a wide range of liquid fractions, as summarized in Fig.~\ref{relative}(b).
The exponent $\alpha_2 = 0.5$ is consistent with the classical Lucas-Washburn imbibition dynamics, demonstrating that the intrinsic liquid transport relative to the foam matrix remains purely capillary-driven.

\section*{Discussion}
We now discuss the physical origin of the apparent super-diffusive imbibition dynamics observed in the half-open geometry.
Here we provide a minimal physical interpretation for why the foam matrix displacement
scales as $x_{\mathrm{bub}} \propto t$, leading to an apparent super-diffusive behavior
of the liquid front.

Since the relative imbibition distance follows $\Delta x \propto t^{1/2}$,
the relative imbibition velocity is given by
\begin{equation}
u_{\mathrm{rel}} = \frac{\partial \Delta x}{\partial t}
\propto \frac{1}{\Delta x}.
\end{equation}
Within the imbibed region, the drag force exerted by the liquid on the foam matrix
can therefore be written as $F_{\mathrm{drag}} \propto u_{\mathrm{rel}}$.
The total driving force acting on the foam matrix is obtained by integrating this drag
over the imbibed length,
\begin{equation}
F_{\mathrm{drive}} = \int_0^{\Delta x} F_{\mathrm{drag}}\,dx
\sim u_{\mathrm{rel}}\,\Delta x \approx \mathrm{const.}
\end{equation}
Balancing this driving force with the dissipative resistance associated with the
translation of the foam matrix, $F_{\mathrm{diss}} \sim \eta\,u_{\mathrm{bub}}$,
leads to a constant matrix velocity, $u_{\mathrm{bub}} = \mathrm{const.}$,
and thus $x_{\mathrm{bub}} \propto t$.

From the Lucas-Washburn fit of the relative imbibition dynamics,
we independently obtain a diffusion constant $D$.
When the matrix displacement is rescaled as $x_{\mathrm{bub}}/d$
and plotted against the dimensionless time $Dt/d^2$,
data obtained for different bubble sizes collapse onto a single master curve
(Fig.~\ref{bubble}(b)).
This collapse indicates that the matrix displacement is governed by the imbibed length
scale rather than by the instantaneous imbibition velocity,
consistent with the above force-balance argument.
In the half-open geometry, the observed penetration distance is given by
$x_{\mathrm{liq}} = \Delta x + x_{\mathrm{bub}}$.
While the relative imbibition $\Delta x$ follows the Lucas--Washburn scaling,
$\Delta x \propto t^{1/2}$,
the additional contribution from matrix translation,
$x_{\mathrm{bub}} \propto t$,
leads to an apparent super-diffusive growth of $x_{\mathrm{liq}}$.
In the fully open geometry, partial backflow toward the injection side
reduces the net matrix displacement,
resulting in a weaker acceleration compared to the half-open case.

A more pronounced modification occurs in the closed geometry.
At $\phi = 3\%$, the exponent approaches $1/2$,
allowing an estimate of the effective diffusivity from the Lucas--Washburn relation.
Using $D = 7.8~\mathrm{mm}^2/\mathrm{s}$ and $r = 12~\mu\mathrm{m}$,
we obtain an effective driving pressure
$\Pi_{\mathrm{eff}} = 103~\mathrm{Pa}$,
nearly an order of magnitude smaller than the theoretical osmotic pressure
$\Pi_{\mathrm{theo}} = 1.53 \times 10^3~\mathrm{Pa}$~\cite{cantat2013}.
This reduction indicates that, when matrix translation is suppressed,
flow-induced stresses are stored as elastic compression,
thereby decreasing the pressure available for imbibition.
Notably, $\Pi_{\mathrm{eff}}$ is comparable to the drainage threshold
measured in vertically oriented foams~\cite{Kaneda2025},
suggesting that it represents the maximum compressive stress
sustainable without inducing bubble rearrangements.

This interpretation clarifies why previous studies predominantly reported
imbibition exponents close to or smaller than $1/2$~\cite{cohen2013,Mensire2015,Tsuritani2021}.
In strongly confined or very dry foams,
the matrix is effectively unable to translate collectively~\cite{Furuta2016,Yanagisawa2021a}.
Under such constraints, flow-structure coupling generates
compressive stresses rather than collective motion,
reducing the effective driving pressure.
The observed $\alpha < 1/2$ therefore reflects mechanical constraint,
not a breakdown of capillary transport.

Finally, the observed size dependence of the imbibition dynamics can be further understood
by making the dependence of the diffusivity $D$ explicit.
Using $D = \Pi_{\mathrm{eff}} r^2/\eta$,
together with the geometric relation $r \sim R\sqrt{\phi}$ for Plateau borders
and the theoretical scaling of the osmotic pressure $\Pi \sim \gamma/R$,
one finds that the dominant contribution leads to $D \propto R$. 
This scaling explains why foams composed of larger bubbles exhibit faster imbibition,
despite common empirical expectations in industrial applications.
The present results therefore highlight that the apparent discrepancy arises
not from anomalous transport, but from the mechanical response of the foam matrix,
which is usually not accounted for in empirical design rules.

\section*{Conclusions}
In this study, we investigated solvent imbibition into a foam under controlled boundary conditions
and demonstrated that the apparent transport dynamics are governed by the mechanical response
of the foam matrix.
By directly measuring both the liquid penetration and the motion of the foam structure,
we showed that the intrinsic imbibition relative to the foam matrix universally follows
the Lucas--Washburn scaling, $\Delta x \propto t^{1/2}$,
independent of boundary conditions, bubble size, and liquid fraction.

In contrast, the apparent imbibition dynamics observed in the laboratory frame
are strongly modified by boundary-controlled matrix motion.
In a half-open geometry, flow-structure coupling induces a collective translation of the foam matrix
at an approximately constant velocity, leading to an apparent super-diffusive growth of the liquid front.
When matrix motion is suppressed, as in the closed geometry,
the same coupling generates compressive elastic stresses that reduce the effective driving pressure
and slow down imbibition.
These results demonstrate that the absence of visible structural rearrangements
does not imply a purely capillary transport process.

Our findings provide a unified interpretation of previous experimental observations,
including reports of sub-diffusive imbibition in strongly confined or dry foams.
More generally, they highlight that solvent transport in soft jammed materials
cannot be characterized by capillary arguments alone,
but must account for the coupling between flow and structural mechanics,
which is critically controlled by boundary conditions. 
Similar couplings between transport and structural dynamics have also been discussed
in the context of other foam-related processes, such as drainage and spreading
under confinement~\cite{Endo2023, Li2024, Li2025}.
The present study suggests that such couplings can play a key role
whenever transport occurs in mechanically constrained soft jammed systems.

The present framework is expected to apply beyond foams,
to a wide class of soft jammed systems such as emulsions, dense colloids,
and biological tissues, where transport processes occur in mechanically constrained environments.

\subsection*{Materials and Methods}

We used a 5.0~wt\% aqueous solution of the ionic surfactant
tetradecyltrimethylammonium bromide (TTAB, purchased from Wako Pure Chemical Industries,
98\% purity).
This concentration is well above the critical micelle concentration
(CMC = 0.12~wt\%)~\cite{Danov2014},
and the surface tension at this concentration was measured to be
$\gamma = 37$~mN/m.
It is well established that foams stabilized by low-molecular-weight surfactants exhibit highly mobile interfaces with minimal interfacial rigidity, making them suitable for studying capillary-driven transport~\cite{Denkov2009}.
The density of the solution is $\rho = 1.03$~g/cm$^3$,
and the viscosity is $\eta = 1.9$~mPa$\cdot$s.

Foams with different bubble sizes were generated using two types of foam dispensers.
Smaller bubbles were produced using a foam dispenser manufactured by Awahour
(Torigoe Plastic Industry Co., Ltd., Japan),
while larger bubbles were generated using a dispenser from Daiso Industries Co., Ltd. (Japan).
Bubble images were acquired using a digital camera.
Bubble interfaces were manually extracted, and the sizes of bubbles in focus were measured.
For each condition, 200 bubbles were randomly selected to determine the size distribution,
and this procedure was repeated ten times.
The mean bubble radius for the smaller-bubble foam was
$R = 0.15$~mm with a standard deviation of 0.037~mm,
while that for the larger-bubble foam was
$R = 0.22$~mm with a standard deviation of 0.082~mm.
The mean three-dimensional liquid fraction $\phi$
was calculated using
\[
\phi = \frac{m}{\rho V_{\mathrm{foam}}},
\]
where $m$ is the mass of the liquid,
$\rho$ is the liquid density,
and $V_{\mathrm{foam}}$ is the foam volume~\cite{Yanagisawa2021a}.
The experimental uncertainty in $\phi$ was estimated to be 0.005.

Image analysis was performed using ImageJ.
To enhance the visibility of liquid penetration, the injected solution was dyed blue using a food-grade dye (Kyoritsu Co., Japan).
For quantitative determination of the penetration distance, the recorded images were separated into RGB channels.
The intensity of the red channel, $I_{\mathrm{red}}$, which is complementary to blue, was normalized by the blue-channel intensity $I_{\mathrm{blue}}$.
The spatial profile of the averaged intensity ratio
$\langle I_{\mathrm{red}}/I_{\mathrm{blue}} \rangle$
exhibits two distinct slopes.
At smaller $x$, the solution has penetrated into the foam,
whereas at larger $x$, the solution has not yet reached.
We define the liquid penetration distance $x_{\mathrm{liq}}$
as the position corresponding to the slope change on the larger-$x$ side.

\section*{Acknowledgements}
N.~Y. was supported by the JSPS Research Fellowship for Young Scientists (20J11840). 
R. K. was supported by JSPS KAKENHI Grant Number 20H01874. 

\section*{Data availablity}
The data that support the findings of this study are available from the corresponding author upon reasonable request.

\section*{AUTHORS CONTRIBUTIONS}
R.~K. conceived the project. K.~T and N.~Y. performed the experiments and K. T. and R. K. analyzed the data. R.~K. wrote the manuscript.

\section*{COMPETING INTERESTS STATEMENT}
The authors declare that they have no competing interests. 

\section*{CORRESPONDENCE}
Correspondence and requests for materials should be addressed to R.~K. (kurita@tmu.ac.jp).

\bibliography{Foam}

\section*{Supplementary}
\setcounter{figure}{0}
\renewcommand{\thefigure}{S\arabic{figure}}

\subsection*{Movie information}
Three movies are provided, corresponding to the fully open, half-open, and closed boundary conditions.
The movies are recorded in real time. The width of the foam region is 30~mm.
The liquid fraction and mean bubble diameter are $\phi = 0.10$ and $d = 0.15$~mm, respectively.
The injected solution is dyed blue with food coloring for clear visualization.

\subsection*{Imbibition dynamics in the fully open geometry}
Figure~S1 shows the time evolution of the liquid penetration distance $x_{\mathrm{liq}}$ and the positions of the left and right edges of the foam matrix in the fully open geometry, where both sides of the foam are open.
In this configuration, the foam matrix is observed to move not only toward the non-injection side but also toward the injection side during imbibition.

\begin{figure}[H]
\begin{center}
\includegraphics[width=150mm]{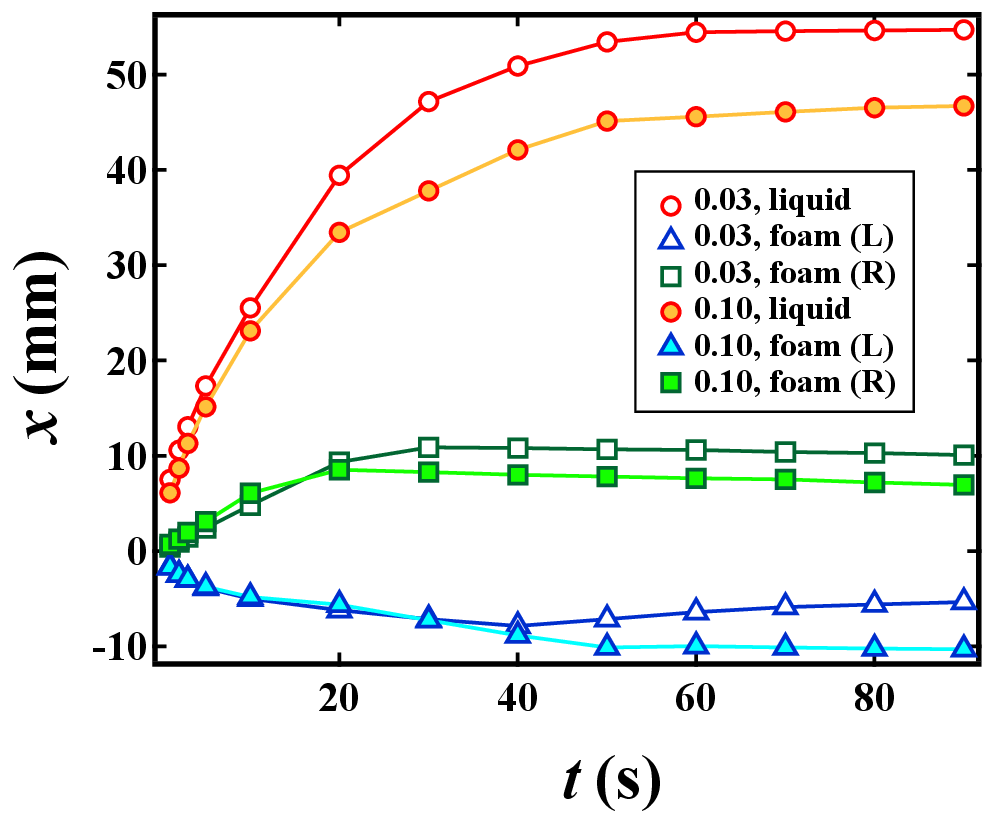}
\end{center}
\caption{
Imbibition dynamics in the fully open geometry.
Time evolution of the liquid penetration distance $x_{\mathrm{liq}}$ and the positions of the left (L) and right (R) edges of the foam matrix for $\phi = 0.03$ and $\phi = 0.10$.
Both edges of the foam matrix move during imbibition, indicating that the foam translates toward both the injection and non-injection sides when both boundaries are open.
}
\label{fig:S1}
\end{figure}

\clearpage

\end{document}